\documentclass[a4paper]{article}
\usepackage{spconf,amsmath,amssymb,bm,graphicx,subcaption}
\usepackage{algorithm}
\usepackage{algorithmic}
\usepackage{xcolor}
\usepackage{bbm}
\usepackage{float}
\ninept
\newfloat{algorithm}{t}{lop}
\graphicspath{{Figures/}}
\newtheorem{definition}{Definition}
\newcommand{\bd}[1]{\bm{#1}}
\title{ Distributed Network Privacy using Error Correcting Codes }
\name{Matt O'Connor and W. Bastiaan Kleijn \thanks{This research was supported by a New Zealand Marsden Fund grant.}}
\address{Victoria University of Wellington}
\begin{document}
%\ninept
%
\maketitle
\begin{abstract}
Most current distributed processing research deals with improving the flexibility and convergence speed of algorithms for networks of finite size with no constraints on information sharing and no concept for expected levels of signal privacy. In this work we investigate the concept of data privacy in unbounded public networks, where linear codes are used to create hard limits on the number of nodes contributing to a distributed task. We accomplish this by wrapping local observations in a linear code and intentionally applying symbol errors prior to transmission. If many nodes join the distributed task, a proportional number of symbol errors are introduced into the code leading to decoding failure if the code's predefined symbol error limit is exceeded.
\end{abstract}
\noindent\textbf{Index Terms}: Distributed, privacy, public networks, linear codes

%!TEX root = ecc_dist_priv.tex

\section{Introduction}
\label{s:intro}

In recent years many distributed algorithms for wireless sensor networks (WSNs) have been developed, primarily stemming from the advances made in small and energy efficient processing units and battery technology. These systems offer exciting potential for use in applications such as seismic monitoring \cite{parker2014distributed} and speech enhancement \cite{taseska2014informed,tavakoli2017distributed}. As concepts such as the Internet of Things (IoT) \cite{cui2016internet} matures, the wireless sensor network will become ubiquitous - nearly any technological device in the near future will have the capability to be a part of the IoT. 

Algorithmic developments in WSNs aim to provide distributed solutions for traditional problems such as acoustic beamforming \cite{markovich2015optimal,o2016distributed,o2014diffusion,sherson2016distributed} and image enhancement \cite{o2017distributed}. Distributed sensors are exploited to collaboratively solve tasks in a manner optimal for the data present, by sharing local observations in an unrestricted manner. The rapidly growing field of distributed optimization \cite{boyd2011distributed} provides a framework for problems of this type and often allows for the computation of distributed solutions that are equal in performance to their centrally computed alternatives. However, these advances are not without their challenges. 

Many devices absorbed into the IoT, or designed as part of large-scale public sensor networks, will have physical sensors, such as microphones or cameras, offering major concerns for the privacy of device owners, users, and the general public. Approaches such as \cite{aono2016privacy} attempt to retain privacy by performing computations in encrypted domains, but this is often computationally expensive, particularly relative to the low compute and low energy world of WSNs. The methods \cite {mo2016privacy,gupta2017privacy} focus on preserving the privacy of node observations by ensuring that other nodes do not have access to these values - only combinations of neighbourhood observations are shared. However, this does not prevent the distributed average from being shared across the entire network. There is still no defence against distant nodes participating in an aggregation task that may enhance a physical signal, thereby compromising the expected privacy of this information.

Most current literature exclusively deals with the case of bounded physical networks and attempts to design algorithms that converge to optimal points given all network data. However, the WSNs considered in this work may operate over many buildings or public spaces, forming practically unbounded networks. The recent advances in graph signal processing and filtering \cite{6494675,6409473,7581108} provide useful tools for applying traditional filtering concepts to graph signals, and would be appropriate for an unbounded paradigm. However, how this may be applied to data privacy has yet to be investigated.

In this work, we consider dealing with effectively unbounded public WSNs where a user may tap the network at any point to designate a query node. The task to be performed is then shared to form a task subnet, and optimized in such a way that distant nodes within the network can neither contribute nor observe information shared with the query node. We aim to enforce hard limits on the distance that data may travel through the network by wrapping messages in linear codes. By applying forced errors to the local initial codes and performing aggregation in the codeword domain, we enable signal enhancement near to the query node while guaranteeing data destruction if the task is shared with a subnet larger than the specified threshold. This allows for a public network with privacy, where many different users may independently access the network for processing while maintaining an expected level of security consistent with the signal type, e.g., it should not be possible to eavesdrop across a large building but it should be possible if you are within the same room.

The following sections are organised as follows: Section \ref{sec:prob_form} summarizes the notation, network setup, and aggregation technique used in this paper; Section \ref{sec:linear_codes} describes linear codes for error correction; Section \ref{sec:ecc_dist_priv} develops an approach for privacy-preserving data summation; Section \ref{sec:simulation} presents simulated experiments on toy data, as well as on the practical task of acoustic beamforming for speech enhancement, confirming the validity of our approach; and finally Section \ref{sec:conclusion} provides some concluding thoughts.

\section{Problem formulation}
\label{sec:prob_form}

In this section we present the notation used throughout the paper, formulate the problem we aim to solve, and summarize existing methods that are currently used to tackle this problem. We discuss that while there are approaches that retain the privacy of individual node messages, the ability for aggregation tasks to spread through a network unchecked has not yet been addressed.

%\subsection{Notation}

An effectively unbounded public wireless sensor network (WSN) may be described as an undirected graph $\mathcal{G} = (\mathcal{V},\mathcal{E})$ consisting of vertices, or nodes, $\mathcal{V}$ connected via edges $\mathcal{E}$. The node set has cardinality $|\mathcal{V}| = V$. Each node is equipped with on-board processors, two-way communication systems, and sensors for a specific signal processing task. We assume that communication in our graph is undirected and contains self-loops. Scalars are denoted with lower case regular font $x$, vectors are boldface lower case $\bd{x}$, while matrices are upper case boldface $\bd{P}$. We use subscripts to designate that a specific variable is owned by a node and superscripts to indicate an update index for iterative algorithms, therefore $\bd{x}_i^k$ describes a vector held by node $i$ at iteration $k$. $[\bd{x}]_i$ denotes the $i$th element of vector $\bd{x}$, $[\bd{P}]_{i,j}$ selects out the scalar entry at row $i$ and column $j$ of matrix $\bd{P}$. We denote selection of multiple elements from a vector as $[\bd{x}]_{\bd{e}}$, where $\bd{e}$ is the vector of indices from which to select, resulting in a vector with dimensionality equal to the number of elements selected. $\bd{x}^T$ denotes vector transpose.

%\subsection{Problem Formulation}

Each node $i\in \mathcal{V}$ holds observation data $\bd{u}_i(t) : \mathbb{Z} \to \mathbb{R}^l$ at time sample $t$. When a user wishes to begin a task, they tap a node that will henceforth be considered the \textit{query node} for that task, denoted by the specific subscript index $q$. For both practical and privacy reasons, the query node spreads the task to a subset $\mathcal{V}_q \subset \mathcal{V}$ of nearby nodes. The goal is for this subset, which forms its own connected subnetwork $\mathcal{G}_q = (\mathcal{V}_q,\mathcal{E}_q)$ with edge set $\mathcal{E}_q \subset \mathcal{E}$, to collaboratively solve the seeded task while also limiting the ability for nodes to join the task that are not within the task subnet. Practically, this smaller task subnet $\mathcal{V}_q$ allows for more efficient computations to be performed since information is not required to propagate through the entire public network. For privacy purposes this reduced information travel distance means that expected levels of privacy are more easily retained - nodes that are very distant from a query node should not have access to tasks seeded at the query node. However, enforcing the size of subnet $\mathcal{V}_q$ is not trivial, particularly if some nodes become compromised and actively wish to spread tasks further than intended.

Formally, we consider an aggregation process where information that has an expected level of privacy, such as an acoustic signal that is assumed to decay with distance as it propagates through air, is combined in a distributed manner as the weighted sum
\begin{equation}\label{eq:step2}
 \sum_{i \in \mathcal{V}_q} a_i \bd{u}_i,
\end{equation}
where $\bd{u}_i$ is observed data at each node $i$, the time index $t$ has been omitted due to the aggregate of each sample being computed independently, and $a_i$ is some scalar. Specific examples of these scalars could be $a_i=1 \; \forall i \in \mathcal{V}_q$, which corresponds to a simple summation, or $a_i=1/V_q \; \forall i \in \mathcal{V}_q$, which would result in a uniform average of observations.

%\subsection{Distributed Summation and Average Consensus}

In order to compute the weighted summation (\ref{eq:step2}) in a distributed manner, each node $i$ may linearly combine data from neighbouring nodes iteratively to produce new estimates. This may be framed generally as
\begin{align}\label{eq:gen_recur}
{\bd{u}}_i^{k+1} &= \sum_{j\in\mathcal{N}_i}[\bd{P}^{k+1}]_{i,j}{\bd{u}}_j^k \quad \forall i \in \mathcal{V}_q
\end{align}
where $\bd{P}^{k+1}$ is a mixing matrix for iteration $k+1$ constrained to have a sparsity pattern according to the topology of the network, such that $\bd{P} \in \mathcal{S}$ where $\mathcal{S} = \{ \bd{P} \in \mathbb{R}^{|\mathcal{V}_q|\times |\mathcal{V}_q|} | \{i,j\}\notin \mathcal{E}_q \Rightarrow [\bd{P}]_{i,j}=0\}$. Note that $\bd{P}^{k+1}$ may change for each iteration, and does not necessarily mix information over all neighbourhoods. Two specific cases of the general mixing (\ref{eq:gen_recur}) frequently seen in the literature, and often useful in practice, are routed protocols \cite{qiu2009enhanced,zhao2004wireless,akkaya2005survey,treerouting} and average consensus \cite{scherber2004locally,xiao2004fast,boyd2005gossip,aysal2008distributed}. Since iterations may be performed at certain nodes before others, routing protocols that remove edges from the query subnet to form a tree topology rooted on the query node may be implemented using (\ref{eq:gen_recur}) by ensuring mixing is performed from leaf to parent nodes. After the routing summation is complete, the query node will have access to the sum of all nodes within the query subnet. In this work, we focus on summations using routed protocols.

To retain privacy when performing iterations such as (\ref{eq:gen_recur}), recent methods \cite{mo2016privacy,gupta2017privacy} aim to perform neighbourhood mixing in such a way that nodes do not directly observe data other than their own. This maintains the privacy of observed signals between nodes, since local observations are not explicitly shared with neighbours, but \textit{does not limit the distance that these mixtures travel within the larger public WSN} $\mathcal{V}$. If, for example, the mixture $\bd{u}_i^{k+1}$ in (\ref{eq:step3}) was an acoustic speech enhancement estimate, then this information would be allowed to travel unboundedly within the network, allowing for distant eavesdropping.

\section{Linear codes over prime fields}
\label{sec:linear_codes}

Linear error correcting codes, or simply linear codes, e.g., \cite{huffman2010fundamentals, imai2014essentials}, are an important class of forward error correcting codes used to protect information transmission or storage from symbol errors by using redundancy. They are defined over a finite vector space $\mathbb{F}_r^n$, where $\mathbb{F}_r$ is a finite field of order $r$. The number of elements $r$ must be a prime power, i.e., $r = p^s$ where $p$ is a prime number and $s$ is a positive integer. In this work we limit ourselves to prime fields, where $s=1 \implies r=p$. The following definition holds:
\begin{definition}\label{def:lin_code}
	(Linear code). A linear code $C$ is a code in $\mathbb{F}_p^n$ for which, whenever $\bd{x},\bd{y} \in C$, then $a\bd{x}+b\bd{y} \in C$, for all $a,b \in \mathbb{F}_p$, i.e., $C$ is a linear subspace of $\mathbb{F}_p^n$.
\end{definition}

A linear code $C$ defines an encoder map $E_{l,n}^p: \mathbb{F}_p^l \to \mathbb{F}_p^n$ from an $l$-dimensional message $\bd{m} \in \mathbb{F}_p^l$ to an $n$-dimensional codeword $\bd{c} \in \mathbb{F}_p^n$. This encoder map is usually computed using the generator matrix $\bd{G}\in \mathbb{F}_p^{l\times n}$, where encoding is performed as
\begin{equation}
\bd{c} = \bd{G} \bd{m}.
\end{equation}
The corresponding decoder map $D_{n,l}^p: \mathbb{F}_p^n \to \mathbb{F}_p^l$ recovers the original message $\bd{m}$ from the codeword $\bd{c}$. Linear codes are typically denoted as $[n,l,d]$ codes, where $n$ is the length of the codeword, $l$ is the length of the message to be encoded, and $d$ refers to the minimum Hamming distance between any two codewords. Every linear code satisfies the Singleton bound $l+d\leq n+1$, where $1 \leq l \leq n$. Given the Hamming distance $d$, a linear code may correctly decode a corrupted codeword $\tilde{\bd{c}}$ provided that fewer than $d/2$ symbol errors occur.

We assume the observation $\bd{u}_i \in \mathbb{R}^l$ at each node $i \in \mathcal{V}$ has entries bounded in the range $-y \preceq \bd{u}_i \preceq y$. Given that the observed data may be continuous (or stored using quantization at a far finer level than our transmission rate would allow, so as to appear effectively continuous), a quantization step may be required prior to coding. The observations are quantized using a uniform $p$-level quantizer $Q_{l,p}$ resulting in $p$ equally spaced values over the range $-y$ to $y$. We refer to the quantized observations as messages $\bd{m}_i = Q_{l,p}(\bd{u}_i) \in \mathbb{F}_p^l \; \forall i \in \mathcal{V}$, and their decoded approximations as $\hat{\bd{u}}_i = R_{l,p}(Q_{l,p}(\bd{u}_i)) \in \mathbb{R}^l \; \forall i \in \mathcal{V}$. In the remainder of this work we refer to $Q$ as the quantizer and $R$ as the dequantizer (rather than as the decoder, to avoid confusion with the linear code decoder $D_{n,l}^p$).

%% old content
\iffalse
Note that although the dequantized values $\hat{\bd{u}}_i$ may only take on a finite number of values, they exist in a vector space over the field of real numbers $\mathbb{R}^l$ rather than the finite vector space in which the messages $\bd{m}_i$ lie.

Therefore, when adding or subtracting codewords $\bd{c}_1 \in \mathbb{F}_r^n$ and $\bd{c}_2 \in \mathbb{F}_r^n$, the operations are performed modulo $p$ and not modulo $r$.

Given that the data of interest may be continuous, a quantization step may be required prior to coding. We denote a general $p$-level uniform quantization mapping for an $m$ dimensional real vector as $Q_{m,p}: \mathbb{R}^m \to \mathbb{F}_p^m$, while we denote the approximate inverse mapping $R_{m,p}: \mathbb{F}_p^m \to \mathbb{R}^m$. 

In this section we describe linear error correcting codes as they appear in the literature. 

\fi
%!TEX root = ecc_dist_priv.tex

\section{Distributed private summation}
\label{sec:ecc_dist_priv}

In this section, we exploit linear codes in a novel way to guarantee distributed privacy when performing processing over a WSN, where a processing task is defined over a subset $\mathcal{V}_q \subset \mathcal{V}$ of nodes originating from a query node $q$. We use linear codes to limit the range that information may travel within the network. We accomplish this by encoding our messages using a linear code and then applying forced errors to the codewords prior to transmission. 

As a result of Definition \ref{def:lin_code}, for any linear code the sum or difference of any two codewords is also a codeword. Addition and subtraction over $\mathbb{F}_p$ are performed modulo the characteristic of the field, which is the prime number $p$. Since messages are encoded by a matrix multiplication with the generator $\bd{G}$, the codeword associated with the sum of two messages $\bd{m}_1$ and $\bd{m}_2$ is the same as the sum of the two separate encodings, i.e.,
\begin{equation}
\begin{aligned}\label{eq:sum_message_codeword}
& E_{l,n}^p(\bd{m}_1 + \bd{m}_2) = \bd{G}(\bd{m}_1 + \bd{m}_2) \\
=& \bd{G}\bd{m}_1 + \bd{G}\bd{m}_2 = E_{l,n}^p(\bd{m}_1) + E_{l,n}^p(\bd{m}_2),
\end{aligned}
\end{equation}
where addition and matrix multiplication are performed using finite field arithmetic over the field $\mathbb{F}_p$.

From (\ref{eq:sum_message_codeword}), performing summations over the network may be performed on \textit{codewords}, rather than messages, with forced errors present. If a node wishes to output an estimate of the private aggregation procedure, only then will it decode the forcibly corrupted codeword mixture. By controlling the number of errors introduced and the Hamming distance of the linear code used, we effectively bound the overall number of nodes able to participate in a distributed task. If more nodes than the limit join a task, then more than $d/2$ errors are introduced resulting in erroneous decodings. For a maximum distance separable (MDS) linear code \cite{huffman2010fundamentals} with Hamming distance $d$, the number of nodes able to join a task without affecting the ability to decode is given by
\begin{equation}
|\mathcal{V}_q| < \frac{d}{2\lambda} = \frac{n-l+1}{2\lambda},
\end{equation}
where $\lambda$ is the number of random errors applied to each node's codeword locally and independently.

Algorithm \ref{alg:priv_summ_prime} describes the Distributed Private Summation (DPS) procedure. For this scenario, we require a message quantizer/dequantizer to map between $\mathbb{R}^l$ and $\mathbb{F}_p^l$, and a linear encoder/decoder to map between $\mathbb{F}_p^l$ and $\mathbb{F}_p^n$.  The parameter $l$ is the dimensionality of the observation. A predefined code length $n$ and a predefined number of symbol errors $\lambda$ are also necessary. Given these requirements, each node determines $\lambda$ codeword symbol indices that will be corrupted by error.

Nodes observe signals $\bd{u}_i \; \forall i \in \mathcal{V}$, and a summation task is defined over a task subnet $\mathcal{V}_q \subset \mathcal{V}$. A set of edges $\mathcal{T}_q^0$ is determined that converts the general task graph into a tree graph rooted at the query node $q$. Messages $\bd{m}_i \; \forall i \in \mathcal{V}$ are formed by mapping observations to the finite field $\bd{F}_p^l$, where message values must satisfy $|\mathcal{V}_q|\mathrm{max}(\|\bd{m}_0\|_\infty,\dots,\|\bd{m}_{|\mathcal{V}_q|}\|_\infty) \leq p$ to guarantee summation overflow does not occur. Initial codewords $\bd{c}_i^0$ are computed by encoding messages $\bd{m}_i, \; \forall i \in \mathcal{V}$, and $\lambda$ symbol errors are applied to each codeword randomly and independently. We then begin iteratively summing through the tree, from leaf nodes to the root. At each iteration $k$ we use the tree edges $\mathcal{T}_q^k$ to define a leaf node set $\mathcal{L}_q^k$, a set of leaf parent nodes $\mathcal{P}_q^k$, and a set of all edges connected to leaf nodes denoted $\mathcal{F}_q^k$. Each leaf parent stores the sum of its own codeword and the codewords of all its leaf neighbours (defined as the union of the leaf parent's neighbours and the current leaf nodes) as $\bd{c}_i^{k+1}$. The tree edge set is then updated by removing the current leaf edge set $\mathcal{F}_q^k$ from the current tree edge set. The final output at the query node $q$ is the decoded and dequantized codeword after summation termination. Note that the total network summation is only available to the query node in this scheme.

\begin{algorithm}
	\caption{Distributed Private Summation}
	\begin{algorithmic}\label{alg:priv_summ_prime}
		\REQUIRE Task subnet $\mathcal{G}_q$; prime field characteristic $p$; message quantizer $Q_{l,p}$; message dequantizer $R_{l,p}$; linear encoder $E_{l,n}^p$; linear decoder $D_{n,l}^p$; code length $n$; number of errors $\lambda$
		\STATE Symbol error index vectors $\bd{e}_i = \bd{a} \sim U(n,\lambda) \; \forall i \in \mathcal{V}$\\[1em]
		\STATE Nodes observe $\bd{u}_i \in \mathbb{R}^l \; \forall i \in \mathcal{V}$
		\STATE Summation defined over task subnet $(\mathcal{V}_q, \mathcal{E}_q) \subseteq (\mathcal{V}, \mathcal{E})$
		\STATE Construct tree edge set $\mathcal{T}_q^0 \subseteq \mathcal{E}_q$ rooted at query node $q$
		\STATE $\bd{m}_i = Q_{l,p}(\bd{u}_i)$, $|\mathcal{V}_q|\mathrm{max}(\|\bd{m}_0\|_\infty,\dots,\|\bd{m}_n\|_\infty) \leq p$
		\STATE Encode $\bd{c}_i^0 = E_{l,n}^p(\bd{m}_i) \in \mathbb{F}_p^n \; \forall i \in \mathcal{V}_q$
		\STATE Apply errors $[\bd{c}_i^0]_{\bd{e}_i} = \bd{b} \sim U(p,\lambda) \; \forall i \in \mathcal{V}_q$
		%\STATE $\bd{c}_q^0 = [\bd{c}_1^T, \bd{c}_2^T, \dots,\bd{c}_{|\mathcal{V}_q|}^T]^T$\\[1em]
		\STATE $k=0$
		\WHILE{$\mathcal{T}_q^k \neq \emptyset$}
		%\FOR{node $i \in \mathcal{V}_q$}
		%\STATE Transmit codeword $\bd{c}_i^k$ to neighbours $j \in \mathcal{N}_i \; \forall i \in \mathcal{V}_q$
		\STATE Define leaf nodes $\mathcal{L}_q^k$, leaf parents $\mathcal{P}_q^k$, and leaf edges $\mathcal{F}_q^k = \{(i,j)|j \in \mathcal{N}_i \; \forall i \in \mathcal{L}_q^k \}$ using  $\mathcal{T}_q^k$
		\STATE $[\bd{P}^{k+1}]_{i,j} = 1$ for $i==j$, 
		\STATE $[\bd{P}^{k+1}]_{i,j} = 1$ for $j\in\mathcal{N}_i \cap \mathcal{L}_q^k \; \forall i \in \mathcal{P}_q^k$,
		\STATE $[\bd{P}^{k+1}]_{i,j} = 0$ otherwise.
		\STATE $\mathcal{T}_q^{k+1} \gets \mathcal{T}_q^k \setminus \mathcal{F}_q^k$\\[0.5em]
		\STATE $\bd{c}_i^{k+1} \gets \sum_{j\in\mathcal{N}_i} [\bd{P}^{k+1}]_{i,j}\bd{c}_j^{k} \; \forall i \in \mathcal{V}_q$
		%\ENDFOR
		\STATE $k \gets k+1$
		\ENDWHILE\\[1em]
		
		\STATE $\bd{u}_q^{\mathrm{Sum}} = R_{l,p}(D_{n,l}^p(\bd{c}_q^k))$
	\end{algorithmic}
\end{algorithm}

\section{Simulated Experiments}
\label{sec:simulation}
In this section we investigate Algorithm \ref{alg:priv_summ_prime} when applied in two simulated scenarios, using Reed-Solomon (RS) codes \cite{reed1960polynomial}. In the first scenario, we seed nodes with random two dimensional message vectors. The mean squared error of the sum, as read at the network query node, is then plotted as a function of nodes contributing to the task. The second scenario applies private summation to distributed audio enhancement, specifically the aggregation step of a delay-and-sum (DSB) and minimum variance distortionless response (MVDR) beamformer. The signal-to-noise ratio of the resulting output is then plotted as a function of task nodes.

The network consists of a varying number of nodes uniformly randomly scattered in a $2D$ circular surface with radius $25$~m. The nodes have a communication range of $10$~m. The binary adjacency matrix for the network graph is then constructed, where connected node pairs are represented with edges of value $1$ while unconnected pairs have edge value $0$.

\subsection{Toy Data}

In this toy scenario, nodes are each assigned a random integer message $\bd{m}_i \in \mathbb{F}_p^l \; \forall i \in \mathcal{V}_q$, where each dimension is drawn from the discrete uniform distribution $\mathcal{U}\{0,p\}$. We use a prime field with characteristic $p=251$ so that our messages may be stored within $8$ bits. A random node is then selected as the query node $q$. We use a fixed codeword length of $n=64$ while varying the message dimensionality, and correspondingly the codeword Hamming distance. Three message lengths of $l=16, 32$, and $48$ are implemented in order to compare the point at which decoding breaks down. This gives a redundancy of $300 \%$, $100 \%$, and $33.3 \%$, respectively. Nodes each introduce one symbol error to their own message prior to beginning the summation procedure. Since the generator matrix $G$ for the RS code used is in standard form, it is possible to na\"ively attempt decoding past the point at which the RS code breaks down by simply reading the first two codeword dimensions. This is used to compute the mean squared error (MSE) after coding fails.

\begin{figure}
	%\vspace{-0.1cm}
	%\hspace{-0.35cm}
	%\vspace{-20cm}
	%\centering
	\includegraphics[width=1\linewidth]{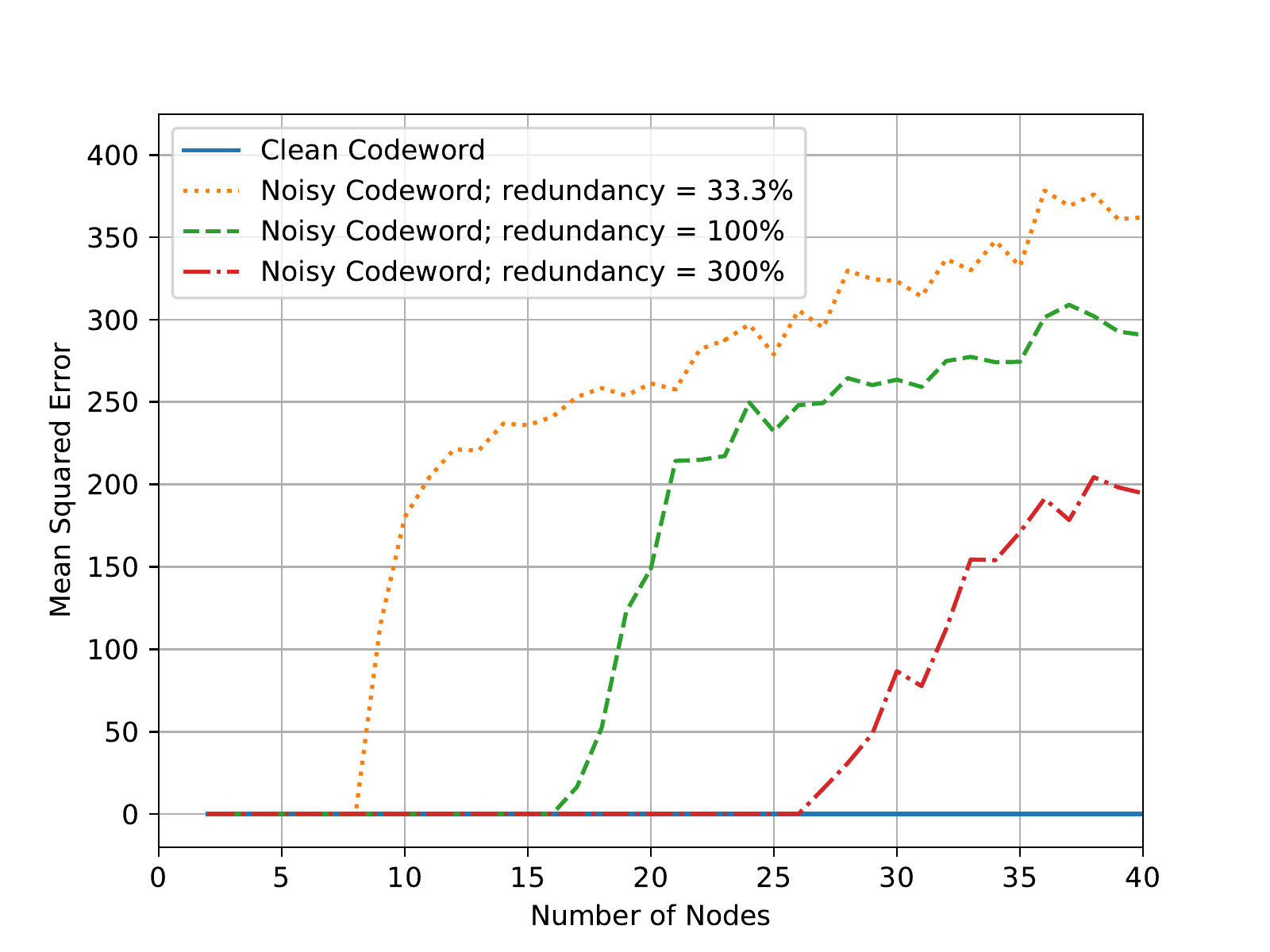}
	\caption{MSE versus number of task nodes for toy data summation, for varying levels of codeword redundancy.}
	\label{fig:mse_vs_distance}
	%\vspace{-0.4cm}
\end{figure}

Figure \ref{fig:mse_vs_distance} plots the MSE of the decoded two dimensional message sum as a function of task nodes. For all code lengths we see decoding error appearing past the expected point of $d/2 = 8$, $16$, and $24$ nodes, respectively, since at this point there are approximately $d/2$ symbol errors in the final summation total.

\subsection{Private Beamforming}

In this private beamforming setup nodes observe acoustic signals originating from a talker located at the centre of the simulated environment surface. A second interfering talker along with independent additive white Gaussian noise at each node is also present. The observed signals at each node, sampled synchronously at $8$~kHz, are calculated using the acoustic transfer function vector $\bd{d}$ computed by assuming a free field model for both talkers. Beamforming is accomplished using an estimated covariance matrix over all nodes $\bd{R}$ to compute the optimal weight vector $\bd{w}^* = (\bd{R}^{-1} \bd{d}) / (\bd{d}^H \bd{R}^{-1}\bd{d})$. The task-specific weighting vector $\bd{w}^*$ may be computed either centrally or in a distributed manner \cite{boyd2011distributed,zhang2017distributed,o2018function}, and we assume that no private data leakage occurs here. For the DSB scenario, the covariance matrix $\bd{R}$ was assumed to be diagonal, while the MVDR beamformer exploited the full matrix.

We process local signals by taking $50$\% overlapping time-domain blocks and applying a Hann window prior to taking the short-time Fourier transform. This gives us frequency-domain signals denoted $\bd{x}_i \in \mathbb{C}^l$ at all nodes $i \in \mathcal{V}_q$ within the query subnet. We denote the stacked collection of these distributed signals as the matrix $\bd{X}_q \in \mathbb{C}^{V_q\times l}$. The complex weight vector $\bd{w}^*\in \mathbb{C}^{|\mathcal{V}_q|}$ may then be used to form signal aggregation across the network as
\begin{equation}\label{eq:step3}
\hat{\bd{s}} = \bd{w}^{*{T}} \bd{X}_q = \sum_{i \in \mathcal{V}_q} [\bd{w}^*]_i \bd{x}_i = \sum_{i \in \mathcal{V}_q} a_i\bd{u}_i,
\end{equation}
where $\bd{u}_i = \mathrm{real}([\bd{w}^*]_i \bd{x}_i) \in \mathbb{R}^l$ is a task weighing at node $i$ applied to the signal at this node to produce observation data $\bd{u}_i$, and $\hat{\bd{s}}$ is the enhanced signal sample. These observations are then quantized to give messages $\bd{m}_i \in \mathbb{F}_p^l \; \forall i \in \mathcal{V}_q$. We use a prime field with characteristic $p=1021$ so that our messages may be stored within $10$ bits. A block size of $l=224$ is used, while the codeword length is $n=255$. This results in a Hamming distance of $d=32$, and a transmission redundancy of $13.8\%$.

\begin{figure}
	%\vspace{-0.1cm}
	%\hspace{-0.35cm}
	%\vspace{-20cm}
	\centering
	\includegraphics[width=1\linewidth]{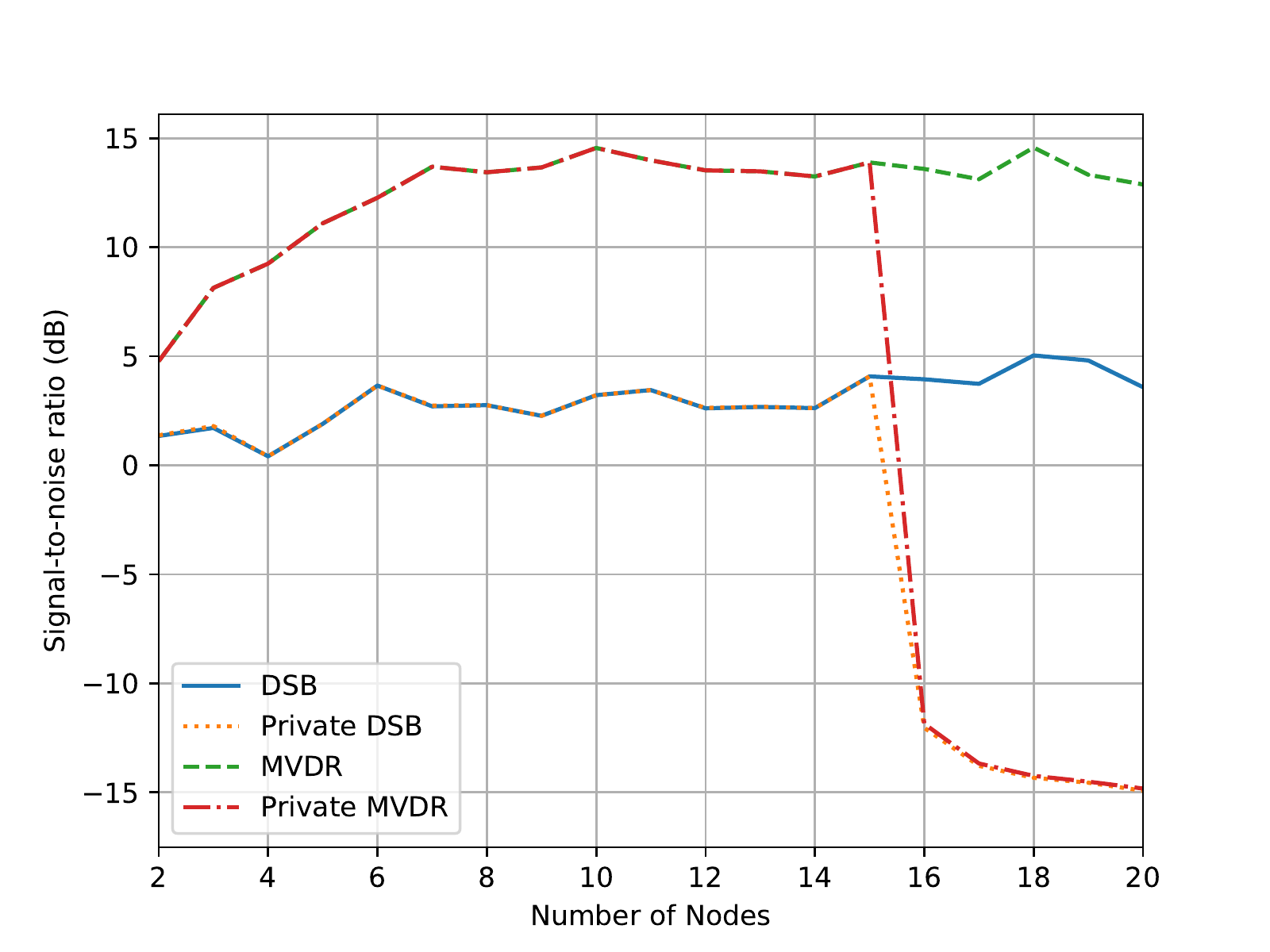}
	\caption{SNR versus number of task nodes for conventional and private beamformers, with codeword redundancy of $13.8\%$.}
	\label{fig:snr_vs_distance}
	%\vspace{-0.4cm}
\end{figure}

Figure \ref{fig:snr_vs_distance} plots the signal-to-noise ratio (SNR) as a function of contributing task nodes. Initially, as more nodes are included in the beamforming task we see an increase in performance of the enhanced signal. This boost in SNR drops after $d/2 = 16$ nodes are included in the task, since at this point decoding fails to output the correct summation total. In contrast we see that with, no errors applied the DSB and MVDR performance continues to rise as more nodes are included, compromising privacy. We note that the point at which information destruction occurs is entirely controlled by the system designer, and depends on the node density of the network (i.e., the number of nodes per meter that the codewords travel through), the Hamming distance of the code used, and the number of errors introduced at each node. This may be set to impose dropoff faster than natural acoustic signal decay, guaranteeing privacy.

\section{Conclusion}
\label{sec:conclusion}

We conclude that public WSN privacy can be ensured by limiting information propagation throughout an unbounded network, where tasks are seeded by user-accessed query nodes. We have applied errors to locally encoded data observations, allowing for distributed aggregation that is performed in a manner that guarantees information destruction when too many nodes contribute to the task. This enforces a level of privacy proportional to the distance a signal travels through the network. Our approach is flexible, scalable, and may be used in combination with other existing protocols that encourage local node privacy.

\vfill\pagebreak
%\newpage\newpage

\bibliographystyle{IEEEbib}
\bibliography{strings,refs}

\end{document}